\documentclass[eqsecnum,noshowpacs,showkeys,nofootinbib,aps]{revtex4}
\usepackage{amsmath,amssymb}
\usepackage{epsfig}
\usepackage{graphicx}
\usepackage{amssymb}
\usepackage{ulem}
\renewcommand{\theequation}{\arabic{equation}}
\def\beq{\begin{equation}}
\def\eeq{\end{equation}}
\def\bea{\begin{eqnarray}}
\def\eea{\end{eqnarray}}
\def\nn{\nonumber}

\def\pa{\partial}

\def\na{\nabla}

\begin{document}
\title{Rough estimates of solar system gravitomagnetic effects in post-Newtonian gravity}
\author{Soon-Tae Hong}
\email{galaxy.mass@gmail.com}
\affiliation{Center for Quantum Spacetime and Department of Physics, Sogang University, Seoul 04107, Korea}
\date{February 7, 2025}
\begin{abstract}
In order to describe properly the gravity interactions including the mass currents, in the gravitomagnetism we construct four Maxwell type gravitational equations which are shown to be analogs of the Maxwell equations in the electromagnetism. Next, exploiting the Maxwell type gravitational equations, 
we explicitly predict the mass magnetic fields for both the isolated system of the spinning Moon orbiting the spinning Earth and that of the Sun and solar system planets orbiting the spinning Sun, whose phenomenological values have not been evaluated in the precedented Newtonian gravity formalisms. In the gravitomagnetism we also phenomenologically investigate the mass magnetic general relativity (GR) forces associated with the mass magnetic fields, to find that they are extremely small but non-vanishing compared to the corresponding mass electric Newtonian forces. Moreover, the directions of the mass magnetic GR forces for the solar system planets except Venus and Uranus are shown to be anti-parallel to those of their mass electric Newtonian forces. Next we investigate the mass magnetic dipole moment related with the B-ring of Saturn, to evaluate $\vec{m}_{M}(Ring)=-1.141\times 10^{4}~{\rm m^{3}~sec^{-1}}~\hat{\omega}$ with $\hat{\omega}$ being the unit vector along the axis direction of the spinning B-ring. The predicted value of $\vec{m}_{M}(Ring)$ is shown to be directly related with the Cassini data on the total mass of the rings of Saturn. 
\end{abstract}
\maketitle

\section{Introduction}
\setcounter{equation}{0}
\renewcommand{\theequation}{\arabic{section}.\arabic{equation}}

In the weak field approximation, the Einstein field equations in general relativity are 
similar to the Maxwell equations of electromagnetism (EM). For a brief formalism of the EM, see Appendix A. 
Next in the post-Newtonian corrections to the gravitational law there 
have been discussions on the frame dragging effects~\cite{ramirez18,everitt15} and the Lense-Thirring effects~\cite{lense18,iorio09}. 
For more details on these effects, see Refs.~\cite{ramirez18,everitt15,lense18,iorio09} and references therein. 
In investigating phenomenology related with spinning compact objects such as the Sun, the solar system planets and the Moon for example, one can effectively exploit the gravitomagnetism possessing the rotational degrees of freedom (DOF). For more details of the gravitomagnetism, see Refs.~\cite{hobson06,ludwig21,speake22,jones23,lasenby23} and references therein.

Recently, we have proposed~\cite{hong23} a new formalism of the modified linear general relativity (MLGR) which uses the vectorial scheme for the mass scalar and mass vector potentials of the gravitational wave (GW). 
To do this, in the MLGR we have employed the Wald approximation~\cite{lasenby23,hong23,wald84}. To be specific, we have found the GW radiation intensity profile possessing a prolate ellipsoid geometry due to the merging binary compact objects source.
Here the radiation lobe is prolate with long axis perpendicular to the orbital plane~\cite{hong23,landau75}. 
To be more specific, at a given radial distance from the binary compact objects, the GW radiation intensity on the revolution axis of the binary compact objects has been shown to be twice that on the equatorial plane~\cite{hong23}. 

On the other hand, the obliquity of Saturn has been known to be too large to have risen during Saturn's formation from a protoplanetary 
disk or from a large impact~\cite{ward04}. The rings of Saturn have been known to appear about 100 
million years ago, based on the estimated strength of satellite ring torques~\cite{goldreich82} and the estimated rate of darkening of the ice-rich 
materials~\cite{zhang17,iess19}. However, it is uncertain how the young rings of Saturn could have formed so recently. 
The Cassini data recently have been exploited to refine estimates of Saturn's moment of inertia. Moreover, it has been 
proposed~\cite{science22} that Saturn previously possessed an additional satellite, called Chrysalis, which caused the obliquity of Saturn to increase through the Neptune resonance. Next destabilization of Chrysalis's orbit about 100 million years ago can then explain the proximity of the system to the resonance and the formation of the young rings through a grazing encounter of Chrysalis with Saturn. Moreover the simulated Chrysalis has been shown to have multiple encounters with Titan which is the largest satellite of Saturn~\cite{science22}. The other Cassini data also have explicitly shown the total mass of the rings of Saturn~\cite{iess19}.

In this paper, we will theoretically investigate the precision astrophysics phenomenology having the gravitomagnetism related with mass $M$ which is the analogue of the EM~\cite{hong23,wald84,jackson,griffiths99em} associated with charge $Q$. Next we will study the mass magnetic fields for the spinning compact objects in the gravitomagnetism. To be more specific, we will evaluate the mass magnetic fields for both the isolated system of the Moon orbiting the spinning Earth and that of the solar system planets orbiting the spinning Sun. Making use of the mass magnetic fields, we will proceed to predict the ensuing Lorentz type mass 
magnetic general relativity (GR) forces for these systems. Next we will construct the mass magnetic dipole moment in the gravitomagnetism to investigate the phenomenology of the B-ring of Saturn.

In Sec. II, we will first construct the formalism of the gravitomagnetism by exploiting the MLGR. 
Next, in this gravitomagnetism we will formulate the mass electric and mass magnetic fields, 
and the corresponding mass electric Newtonian force and 
mass magnetic one. In Sec. III, we will investigate the phenomenology associated with the mass magnetic fields 
of both the spinning Moon orbiting the spinning Earth, 
and the spinning solar system planets orbiting the spinning Sun. 
Moreover we will evaluate the mass electric Newtonian forces and mass magnetic ones for these systems. 
Next we will evaluate the mass magnetic dipole moment associated with the B-ring of Saturn.
Sec. IV includes conclusions.

\section{Formalism of astrophysical phenomenology}
\setcounter{equation}{0}
\renewcommand{\theequation}{\arabic{section}.\arabic{equation}}
\label{setupgravitosectionph}

\subsection{gravitomagnetism originated from MLGR}

In this subsection, we construct the formalism of the gravitomagnetism by using the MLGR~\cite{hong23}. To do this, we will briefly 
recapitulate the MLGR originated from the Einstein GR. Now 
we start with the Einstein-Hilbert action
\beq
S_{EH}=\int d^{4}x \sqrt{-g}R 
\label{einhil}
\eeq
from which we construct the Einstein GR field equation~\cite{wald84,hobson06,carroll96} 
\beq
R_{\alpha\beta}-\frac{1}{2}g_{\alpha\beta}R=\frac{8\pi G}{c^{4}} T_{\alpha\beta},
\label{eingrf}
\eeq
where $R_{\alpha\beta}$ and $R$ are the Riemann tensor and scalar curvature, respectively, and $T_{\alpha\beta}$ 
is the energy stress tensor. Next we assume that the deviation $h_{\alpha\beta}$ 
of the spacetime metric $g_{\alpha\beta}$ from a flat metric $\eta_{\alpha\beta}$ is small
\beq
g_{\alpha\beta}=\eta_{\alpha\beta}+h_{\alpha\beta}.
\label{gab}
\eeq 
Exploiting the Einstein GR field equation in (\ref{eingrf}) and $g_{\alpha\beta}$ in (\ref{gab}), we construct 
the linearized Einstein equation~\cite{wald84,hong23,hobson06,carroll96}
\beq
-\frac{1}{2}\square \bar{h}_{\alpha\beta}+\pa^{\gamma}\pa_{(\alpha}\bar{h}_{\beta)\gamma}-\frac{1}{2}\eta_{\alpha\beta}
\pa^{\gamma}\pa^{\delta}\bar{h}_{\gamma\delta}=\frac{8\pi G}{c^{4}} T_{\alpha\beta},
\label{linein}
\eeq
where $\square=\pa^{\beta}\pa_{\beta}$ with the metric $\eta_{\alpha\beta}={\rm diag}(-1,+1,+1,+1)$. 
Here the trace reversed perturbation $\bar{h}_{\alpha\beta}$ is given by $\bar{h}_{\alpha\beta}=h_{\alpha\beta}-\frac{1}{2}h$ 
with $h=\eta^{\alpha\beta}\bar{h}_{\alpha\beta}$. Next exploiting (\ref{linein}) and the Lorentz type gauge condition in gravity
\beq
\pa^{\beta}\bar{h}_{\alpha\beta}=0,
\label{pabarh0}
\eeq
we arrive at the Maxwell type equation in the MLGR
\beq
\square \bar{h}_{\alpha\beta}=-\frac{16\pi G}{c^{4}} T_{\alpha\beta}.
\label{squarebarh}
\eeq

Now we consider the non-vanishing $\bar{h}_{00}$ and $\bar{h}_{0i}$ $(i=1,2,3)$, together with 
$\bar{h}_{ij}=0$~\cite{lasenby23,hong23,wald84}. We define $\bar{h}_{00}$ and 
$T_{00}$ as
\beq
\bar{h}_{00}=-\frac{4}{c^{2}}\phi_{M},~~~T_{00}=c^{2}\rho_{M},
\label{barht0}
\eeq
where $\phi_{M}\equiv cA_{M}^{0}$ is the linearized physical quantity in the MLGR. From now on, 
the subscripts $M$ an $Q$ denote the mass and charge physical quantities respectively. 
Inserting (\ref{barht0}) into the Maxwell type equation in (\ref{squarebarh}) yields
\beq
\square \phi_{M}=4\pi G\rho_{M}.
\label{naphi0a}
\eeq
Next we define $\bar{h}_{0i}$ and $T_{0i}$ as
\beq
\bar{h}_{0i}=\frac{4}{c}A^{i}_{M},~~~T_{0i}=-c\rho_{M}v^{i}=-cJ^{i}_{M},
\label{barht2}
\eeq
where the energy stress tensor is approximated to linear order in velocity in the Wald Approximation~\cite{lasenby23,hong23,wald84}. 
In the MLGR, inserting (\ref{barht2}) 
into the Maxwell type equation in (\ref{squarebarh}) produces
\beq
\square A^{i}_{M}=\frac{4\pi G}{c^{2}} J^{i}_{M}.
\label{naphi}
\eeq
Note that in the Wald approximation associated with $T_{00}$ in (\ref{barht0}) and $T_{0i}$ in (\ref{barht2}), 
we construct $T_{\alpha\beta}$ in terms of $J^{i}_{M}=\rho_{M} v^{i}$~\cite{lasenby23,hong23,wald84}
\beq
T_{\alpha\beta}=\left(\begin{array}{cc}
c^{2}\rho_{M} &-cJ^{j}_{M}\\
-cJ^{i}_{M} &0
\end{array}
\right),
\label{energystresstensor2}
\eeq
where the space-space components $T_{ij}$ vanish.

In vacuum, inserting $T_{\alpha\beta}=0$ into (\ref{squarebarh}) and exploiting the definitions of $\bar{h}_{00}$ and $\bar{h}_{0i}$ in (\ref{barht0}) and (\ref{barht2}), we obtain the wave equations for the non-vanishing 
fields $\bar{h}_{00}$ and $\bar{h}_{0i}$
\beq
\square \bar{h}_{00}=\square A_{M}^{0}=0,~~~\square \bar{h}_{0i}=\square A_{M}^{i}=0,
\label{squarebarh4}
\eeq
from which we arrive at the wave equations in the MLGR
\beq
\square A_{M}^{\alpha}=0,~~~\alpha=0,1,2,3
\label{squarebarh5}
\eeq
which describes a massless spin-one graviton propagating in the flat spacetime. This phenomenology in the gravitomagnetism is analogous to 
that in the EM. Note also that we have the Lorentz type gauge condition for $A_{M}^{\alpha}$
\beq
\pa_{\alpha}A_{M}^{\alpha}=0.
\label{pabarh2a}
\eeq

Next we investigate the formalism of the gravitomagnetism. To this end, we define the {\it mass electric and mass magnetic fields} $E_{M}^{i}$ and $B_{M}^{i}$ in terms of $\bar{h}_{00}$ in (\ref{barht0}) and $\bar{h}_{0i}$ in (\ref{barht2}) constructed 
in the relativistic MLGR 
\beq
E_{M}^{i}=-\pa_{i}\phi_{M}-\frac{\pa A_{M}^{i}}{\pa t}=\frac{c^{2}}{4}(\pa_{i}\bar{h}_{00}-\pa_{0}\bar{h}_{0i}),~~~
B_{M}^{i}=\epsilon^{ijk}\pa_{j}A_{M}^{k}=\frac{c}{4}\epsilon^{ijk}\pa_{j}\bar{h}_{0k}.
\label{linem2}
\eeq
Note that $E_{M}^{i}$ contains the DOF related with the term $\frac{\pa A_{M}^{i}}{\pa t}$. 
After some algebra exploiting $\vec{E}_{M}$ and $\vec{B}_{M}$ in (\ref{linem2}) in the MLGR, we construct the Maxwell type equations in the gravitomagnetism~\cite{ludwig21}
\bea
\na\cdot\vec{E}_{M}&=&-4\pi G\rho_{M},~~~\na\times\vec{E}_{M}+\frac{\pa\vec{B}_{M}}{\pa t}=0,\nn\\
\na\cdot\vec{B}_{M}&=&0,~~~~~~~~~~~~~~
\na\times\vec{B}_{M}-\frac{1}{c^{2}}\frac{\pa\vec{E}_{M}}{\pa t}=-\frac{4\pi G}{c^{2}}\rho_{M}\vec{v}.
\label{eqnseightlina}
\eea
Note that the Maxwell type gravitational equations in 
(\ref{eqnseightlina}) in the gravitomagnetism are the analogs of the Maxwell equations in the EM~\cite{jackson,griffiths99em}. 
Note also that in the gravitomagnetism the equations in (\ref{eqnseightlina}) 
include the time derivative terms related with the dynamics of the gravitomagnetism interactions.

Now it seems appropriate to address comments on the gravitomagnetism 
in Refs.~\cite{hobson06,speake22,jones23,lasenby23}, where the gravitational Maxwell equations are given by
\bea
\na\cdot\vec{E}_{g}&=&-4\pi G\rho_{M},~~~\na\times\vec{E}_{g}=0,\nn\\
\na\cdot\vec{B}_{g}&=&0,~~~~~~~~~~~~~~
\na\times\vec{B}_{g}=-\frac{16\pi G}{c^{2}}\rho_{M}\vec{v}.
\label{eqnseightlinagrv}
\eea
First, in the gravitomagnetism in (\ref{eqnseightlinagrv}), we find the curl equations which do not possess the dynamic DOF associated with the time derivative terms in (\ref{eqnseightlina}). Note that the equations for $\vec{E}_{g}$ denote the gravitational field produced by a static mass configuration. Next the equations for $\vec{B}_{g}$ yield a notationally means of determining the extra gravitational field produced by 
moving masses associated with $\rho_{M}\vec{v}$~\cite{hobson06}.

Second, even in the static case, the last equation in (\ref{eqnseightlinagrv}) has a {\it coefficient different} from that in the gravitomagnetism in (\ref{eqnseightlina}). This feature in the gravitomagnetism in 
Refs.~\cite{hobson06,speake22,jones23,lasenby23} originates from the 
identification~\cite{hobson06}:
\beq
\epsilon_{0}\leftrightarrow -\frac{1}{4\pi G},~~~\mu_{0}\leftrightarrow -\frac{16\pi G}{c^{2}}.
\label{rlarrow}
\eeq 
In contrast, in the gravitomagnetism associated with (\ref{eqnseightlina}), we construct the last equation possessing 
$-\frac{4\pi G}{c^{2}}$, by using the treatment of the Einstein GR field equation in 
(\ref{eingrf}) and $g_{\alpha\beta}=\eta_{\alpha\beta}+h_{\alpha\beta}$ in (\ref{gab}) used in the MLGR. Especially in order to predict the astrophysical quantities, which will be {\it numerically} 
evaluated in Section 3 for instance, we need to exploit the equations in (\ref{eqnseightlina}). This 
feature in the MLGR and the ensuing gravitomagnetism associated with (\ref{eqnseightlina}) is one of the main points of this papaer.

Next we have comments on the gauge invariance of the gravitomagnetism and that of 
the EM described in Appendix A. First, in the MLGR~\cite{hong23}, we do not exploit the traceless transverse 
(TT) gauge~\cite{wald84,hobson06,carroll96,hong23} 
which is a second-rank tensorial formalism in the linearized general relativity (LGR). In the TT gauge we choose $\bar{h}_{TT}^{0i}\equiv 0$. Next, for the MLGR, exploiting the Wald approximation~\cite{lasenby23,hong23,wald84}, we have constructed a vectorial formalism which is similar to that of the EM~\cite{hong23}. Note that 
in the MLGR we have the vanishing $\bar{h}^{ij}=0$ associated with the Wald approximation, while in the LGR we possess 
the nonzero $\bar{h}_{TT}^{ij}\neq 0$ in the TT gauge~\cite{wald84,hobson06,carroll96,hong23}.

Second, the equations of motion (EOM) for the fields $\phi_{M}$ and $\vec{A}_{M}$ in 
(\ref{naphi0a}) and (\ref{naphi}) produce the covariant EOM of $A^{\alpha}_{M}\equiv(\frac{\phi_{M}}{c},A^{i}_{M})$
\beq
\square A^{\alpha}_{M}=\frac{4\pi G}{c^{2}} J^{\alpha}_{M},
\label{naphi2}
\eeq
where $J^{\alpha}_{M}\equiv (c\rho_{M},J^{i}_{M})$. Next, (\ref{naphi2}) is the same form of the EOM for the fields $A^{\alpha}_{Q}\equiv(\frac{\phi_{Q}}{c},A^{i}_{Q})$ obtainable from (\ref{naphi0q}) and $J^{\alpha}_{Q}\equiv (c\rho_{Q},J^{i}_{Q})$:
\beq
\square A^{\alpha}_{Q}=-\frac{1}{c^{2}\epsilon_{0}} J^{\alpha}_{Q},
\label{naphi22}
\eeq
which are the covariant EOM. Note that, in the MLGR, the GW radiation intensity profile possess a prolate ellipsoid geometry possessing the angle dependence of the merging binary compact objects source~\cite{hong23}.

Third, motivated by the mathematical similarity in the vectorial forms of the EOM for 
$A_{M}^{\alpha}(=-\frac{c}{4}\bar{h}^{0\alpha})$ ($\alpha=0,1,2,3$) in (\ref{naphi2}) and $A_{Q}^{\alpha}$ in (\ref{naphi22}), 
we construct the vectorial formalism for the spin-one graviton in the MLGR~\cite{hong23}. Now we investigate the gauge invariance and U(1) transformation in the gravitomagnetism. To do this, we start with the Dirac equation for the electron wave function $\psi$ and the mass scalar and mass vector potentials $A_{M}^{\mu}$ in the unit of $\hbar=c=1$ 
\beq
i\gamma^{\mu}\pa_{\mu}\psi-m_{e}\psi-e\gamma^{\mu}A_{M,\mu}\psi=0,
\label{diraceqn}
\eeq
with $m_{e}$ and $e$ being the electron mass and charge, respectively. Next, keeping in mind that in the MLGR we have the spin-one graviton, we study the quantum gravitomagnetic dynamics (QGD) which describes the interactions between electrons and gravitons. Now we introduce the QGD Lagrangian which is analogous to that of the quantum electrodynamics (QED) in (\ref{qcdlag})~\cite{smith86}
\beq
{\cal L}_{QGD}=-\frac{1}{4}F_{M,\mu\nu}F_{M}^{\mu\nu}+\bar{\psi}(i\gamma^{\mu}\pa_{\mu}-m_{e})\psi-e\bar{\psi}\gamma^{\mu}A_{M,\mu}\psi,
\label{qgdlag}
\eeq
where $F_{M}^{\mu\nu}=\pa^{\mu}A_{M}^{\nu}-\pa^{\nu}A_{M}^{\mu}$ and the third term denotes the interaction between the electron and the gravitomagnetic wave (or spin-one graviton). Now we find that the QGD Lagrangian in (\ref{qgdlag}) is invariant under the gauge transformations
\bea
\psi(x)&\rightarrow& \psi^{\prime}(x)=e^{-ie\Theta(x)}\psi(x),\nn\\
A_{M}^{\mu}(x)&\rightarrow& A_{M}^{\prime\mu}(x)=A_{M}^{\mu}(x)+\pa^{\mu}\Theta(x),
\label{trfm2}
\eea
where the first relation is the U(1) transformation.

\subsection{Mass magnetic GR force}

Now we construct the mass magnetic GR force by exploiting the gravitomagnetism which is formulated in the previous subsection.
For the cases of {\it mass electrostatics and mass magnetostatics} which are practically applicable to the phenomenological predictions 
in the next section, the mass electric and mass magnetic fields are given by 
\beq
\vec{E}_{M}=-G\int\frac{\rho_{M}\hat{R}}{R^{2}}d^{3}x^{\prime},~~~
\vec{B}_{M}=-\frac{G}{c^{2}}\int\frac{\vec{J}_{M}\times\hat{R}}{R^{2}}d^{3}x^{\prime},
\label{massemfields}
\eeq
where $\vec{R}=\vec{x}-\vec{x}^{\prime}$ is the vector from $d^{3}x^{\prime}$ to the field point $\vec{x}$ and 
$\hat{R}=\vec{R}/R$ with $R=|\vec{x}-\vec{x}^{\prime}|$. Using (\ref{massemfields}) 
we obtain the corresponding force acting on a test mass $M_{0}$ moving with velocity $v$ as follows\footnote{The superscripts 
$E$ and $B$ stand for the mass {\it electric} Newtonian force and the mass {\it magnetic} force, respectively.}
\beq
\vec{F}_{M}=M_{0}(\vec{E}_{M}+\vec{v}\times\vec{B}_{M})\equiv \vec{F}_{M}^{E}+\vec{F}_{M}^{B}.
\label{calftwo2q0}
\eeq
Now, in the mass magnetostatics, we construct the mass vector potential $\vec{A}_{M}$ as 
\beq
\vec{A}_{M}=-\frac{G}{c^{2}}\int \frac{\vec{J}_{M}}{R}d^{3}x^{\prime}.
\label{vecam}
\eeq
Note that, exploiting $\vec{B}_{M}(=\na\times \vec{A}_{M})$ we find that $\vec{A}_{M}$ in (\ref{vecam}) reproduces 
$\vec{B}_{M}$ in (\ref{massemfields}).

Now it seems appropriate to address some comments on the applications of the formulas (\ref{massemfields}) and (\ref{vecam}). 
First, exploiting (\ref{massemfields}), we construct the mass magnetic field $\vec{B}_{M}$ at the center of the loop of orbital radius $R$ 
which consists of total mass $M$ and orbits its center with orbital period $T$ 
\beq
\vec{B}_{M}=-\frac{2\pi GM}{c^{2}RT}\hat{\omega}.
\label{bmrt}
\eeq   
Here $\hat{\omega}$ is the unit vector along the axis of the loop orbiting its center. Note that, inside (outside) 
the loop, the direction of $\vec{B}_{M}$ is anti-parallel (parallel) to $\hat{\omega}$.

\begin{figure}[t]
\centering
\includegraphics[width=8.0cm]{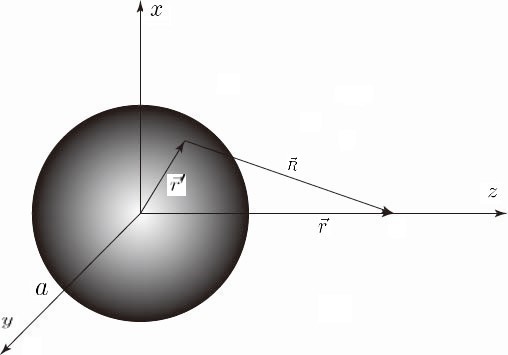}
\caption[hyper] {The geometry for a solid sphere of mass 
$M$ and radius $a$ spinning with angular velocity $\vec{\omega}=\omega\hat{\omega}$ where $\hat{\omega}$ is 
the unit vector along $\hat{x}$. Here $\vec{r}^{\prime}=r^{\prime}(\hat{x}\sin\theta^{\prime}\cos\phi^{\prime}
+\hat{y}\sin\theta^{\prime}\sin\phi^{\prime}+\hat{z}\cos\theta^{\prime})$ and 
$R=(r^{2}+r^{\prime 2}-2rr^{\prime}\cos\theta^{\prime})^{1/2}$ with $\theta^{\prime}$ being the polar angle between $\vec{r}$ 
and $\vec{r}^{\prime}$.} 
\label{solidsp}
\end{figure}

Second, we consider a solid sphere of mass $M$ and radius $a$ spinning with angular velocity $\vec{\omega}=\omega\hat{\omega}$ as 
shown in Figure~\ref{solidsp}. Note that $\vec{J}_{M}=\rho_{M}\vec{v}=\rho_{M}\vec{\omega}\times \vec{r}^{\prime}$ produces
\beq
\vec{J}_{M}=\frac{3M}{4\pi a^{3}}(-\hat{y}~\omega r^{\prime}\cos\theta^{\prime}
+\hat{z}~\omega r^{\prime}\sin\theta^{\prime}\sin\phi^{\prime}).
\label{vecjm1}
\eeq
Inserting (\ref{vecjm1}) into (\ref{vecam}), we find the mass vector potential $\vec{A}_{M}$ as follows
\beq
\vec{A}_{M}=-\frac{3GM\omega}{2c^{2}a^{3}}\int_{0}^{a}dr^{\prime}r^{\prime 3}\int_{0}^{\pi}
\frac{d\theta^{\prime}\sin\theta^{\prime}\cos\theta^{\prime}}{R}\hat{\omega}\times\hat{r},
\label{amvec}
\eeq
where we have used the relation $\hat{y}=-\hat{\omega}\times\hat{z}$. 
Next, on the equatorial plane of the solid sphere, we arrive at $\vec{A}_{M}$ at $\vec{r}=r\hat{r}$ with $r$ $(r\ge a)$
\footnote{Here we have exploited the identity 
$\int_{0}^{\pi}\frac{d\theta^{\prime}\sin\theta^{\prime}\cos\theta^{\prime}}{R}=\frac{2r^{\prime}}{3r^{2}}$.}
\beq
\vec{A}_{M}=-\frac{GMa^{2}\omega}{5c^{2}r^{2}}\hat{\omega}\times\hat{r}.
\label{amph}
\eeq
Exploiting (\ref{amph}), we finally formulate the mass 
magnetic field $\vec{B}_{M}(=\na\times\vec{A}_{M})$ of the form
\beq
\vec{B}_{M}=\frac{4\pi GMa^{2}}{5c^{2}r^{3}t}\hat{\omega},
\label{bmgma2}
\eeq
where $t$ is the rotational period of the spinning solid sphere.
Note that, outside the spinning solid sphere, the direction of $\vec{B}_{M}$ is parallel to $\hat{\omega}$ on the 
equatorial plane. Note also that in the EM the direction of $\vec{B}_{Q}$ circles around the charge current following the right hand thumb rule since the charges interact repulsively~\cite{jackson,griffiths99em}. In contrast, in the gravitomagnetism the 
direction of $\vec{B}_{M}$ circles around the mass current following the left hand thumb rule due to the fact that the masses interact attractively. Explicitly these features can be explained by the comparison of (\ref{vecam}) and (\ref{vecam0}).

\section{Phenomenology of astrophysical systems in gravitomagnetism}
\setcounter{equation}{0}
\renewcommand{\theequation}{\arabic{section}.\arabic{equation}}
\label{setupgravitosectionph}

\subsection{Earth and Moon}

Now in the gravitomagnetism we consider the {\it isolated two body system} of the Earth of mass $M_{\oplus}$ and the spinning Moon of mass $M_{m}$ orbiting the spinning Earth to find the phenomenological prediction for the mass magnetic field 
$\vec{B}_{M}(M_{\oplus})=B_{M}(M_{\oplus})\hat{\omega}$ at the surface of the Earth where 
$\hat{\omega}$ is the unit vector along the axis direction of the spinning Earth. Note that $\hat{\omega}$ is also the unit vector along 
both the axis direction of the spinning Moon and the orbital axis direction of the Moon orbiting the Earth. The mass magnetic field at the surface of the Moon will be considered later. 
Exploiting (\ref{bmrt}) and (\ref{bmgma2}), we construct the expression of mass magnetic field $B_{M}(M_{\oplus})$ for the 
isolated system of the Earth and Moon, both of which possess the rotational DOF, for an observer located on the surface of the 
Earth\footnote{Here we have assumed that this observer resides on a space fixed frame and does not co-rotate together with the Earth so that 
we can include the spinning effect of the Earth. The same statement will be applied to the observers who are located on the surface of the Moon, the Sun and the solar system planets.}
\bea
B_{M}(M_{\oplus})&=&\frac{4\pi GM_{\oplus}}{5c^{2}r_{\oplus}t_{\oplus}}-\frac{2\pi G M_{m}}{c^{2}R_{m}T_{m}}
+\frac{4\pi GM_{m}r_{m}^{2}}{5c^{2}R_{m}^{3}t_{m}}\nn\\
&=&(2.029\times 10^{-14}-3.777\times 10^{-19}+3.085\times 10^{-24})~{\rm sec^{-1}},
\label{bmearth}
\eea
where the first, second and third  terms originate from the contributions of the spinning effect of the Earth and 
orbital effect of the Moon and spinning effect of the Moon, respectively. Here the subscript $m$ denotes the physical quantities of the Moon. 
Next $r_{\oplus}$ and $t_{\oplus}$ are the radius and rotational period of the spinning Earth, 
while $r_{m}$, $R_{m}$, $t_{m}$ and $T_{m}$ are the radius, average orbital distance, rotational period and average orbital period 
of the spinning Moon orbiting the Earth, respectively.\footnote{For the observational data for 
$M_{I}$, $r_{I}$, $R_{I}$, $t_{I}$ and $T_{I}$ $(I=m, \oplus)$ we exploit~\cite{carroll96,pdg,padman}. Similarly, we will 
also use ~\cite{carroll96,pdg,padman} for the 
observational data for $M_{i}$, $r_{i}$, $R_{i}$, $t_{i}$ and $T_{i}$ $(i=1,2,...8)$ for the $i$-th planets in the solar system which will be discussed in the next subsections.} Note that the contributions from the orbital effect of Moon and spinning effect of Moon are negligible. 
The contributions of the other solar system planets to $B_{M}(M_{\oplus})$ will be discussed in (\ref{bmmoplus}).

Next we evaluate the mass magnetic field $\vec{B}_{M}(M_{m})=B_{M}(M_{m})\hat{\omega}$ for an observer located on 
the surface of the Moon by using (\ref{bmgma2}). To do this, we obtain the expression of the mass magnetic field $B_{M}(M_{m})$ 
for the isolated system of the Earth and Moon
\beq
B_{M}(M_{m})=\frac{4\pi GM_{m}}{5c^{2}r_{m}t_{m}}+\frac{4\pi G M_{\oplus}r_{\oplus}^{2}}{5c^{2}R_{m}^{3}t_{\oplus}}
=(3.343\times 10^{-17}+9.267\times 10^{-20})~{\rm sec^{-1}},
\label{bmearth2}
\eeq 
where the first and second terms originate from the contributions of the spinning Moon and the spinning Earth, respectively.

Now we investigate the force acting on the spinning Moon of mass $M_{m}$ due to the spinning Earth of mass $M_{\oplus}$. 
Exploiting $\vec{F}_{M}^{B}$ in (\ref{calftwo2q0}), we are left with $\vec{F}_{M}^{B}(M_{m}-M_{\oplus})=F_{M}^{B}(M_{m}-M_{\oplus})\hat{r}$ with $\hat{r}$ being 
the radial direction from the Earth\footnote{From now on $\vec{F}_{M}^{B}(M_{m}-M_{\oplus})$ denotes that the 
force $\vec{F}_{M}^{B}$ acts on the position at $M_{m}$ under the field at $M_{\oplus}$ for instance. 
Here the argument $M_{m}-M_{\oplus}$ indicates the vector of the direction from $M_{\oplus}$ to $M_{m}$ with magnitude $|M_{m}-M_{\oplus}|$.}
\beq
F_{M}^{B}(M_{m}-M_{\oplus})=(2.513\times 10^{9}+6.966\times 10^{6})~{\rm N}.
\label{fmbmoon0}
\eeq

Next, making use of $\vec{E}_{M}$ in (\ref{massemfields}) and $\vec{F}_{M}^{E}$ in (\ref{calftwo2q0}), one can readily construct 
the mass electric Newtonian force acting on the Moon $\vec{F}_{M}^{E}(M_{m}-M_{\oplus})=F_{M}^{E}(M_{m}-M_{\oplus})\hat{r}$ 
for the isolated system of the Earth and Moon, for an observer located on the surface of the Moon\footnote{From now on the minus sign in $F_{M}^{E}(M_{m}-M_{\oplus})$ indicates the attractive Newtonian force between $M_{m}$ and $M_{\oplus}$ for example.} 
\beq
F_{M}^{E}(M_{m}-M_{\oplus})=-1.983\times 10^{20}~{\rm N},
\label{fMEMoon0}
\eeq
from which we arrive at the characteristic ratio $\chi(M_{m}-M_{\oplus})$ of the form
\beq
\chi(M_{m}-M_{\oplus})=|\vec{F}_{M}^{B}(M_{m}-M_{\oplus})|/|\vec{F}_{M}^{E}(M_{m}-M_{\oplus})|=1.271\times 10^{-11}.
\label{ratioofff}
\eeq
The ratio $\chi(M_{m}-M_{\oplus})$ in (\ref{ratioofff}) implies that the gravitomagnetic correction, namely $|\vec{F}_{M}^{B}(M_{m}-M_{\oplus})|$ 
in the gravitomagnetism, to the Newtonian classical prediction 
$|\vec{F}_{M}^{E}(M_{m}-M_{\oplus})|$ is extremely small but it is non-vanishing.

\subsection{Sun and solar system planets}

In the gravitomagnetism we now investigate the isolated system of the Sun and the solar system planets orbiting the spinning Sun to evaluate the mass magnetic field 
$\vec{B}_{M}(M_{\odot})=B_{M}(M_{\odot})\hat{\omega}$ for an observer located on the surface of the Sun where 
$\hat{\omega}$ is the unit vector along the axis direction of the spinning Sun.

Here, for simplicity, we assume that the rotation 
axes of the solar system planets, except Venus and Uranus, are approximately parallel to that of the Sun to produce the vanishing obliquities 
and that thse axes reside on the equatorial plane of the Sun. 
We also assume that on the equatorial plane of the Sun, the rotation axes of Venus and Uranus are anti-parallel to and perpendicular to that of the Sun, respectively.

Similar to (\ref{bmearth}), exploiting (\ref{bmrt}) and (\ref{bmgma2}) we find the formula of mass magnetic field $B_{M}(M_{\odot})$ 
for the isolated system of the Sun and solar system planets, for an observer located on the surface of the Sun,
\bea
B_{M}(M_{\odot})&=&\frac{4\pi GM_{\odot}}{5c^{2}r_{\odot}t_{\odot}}
-\sum_{i=1}^{8}\frac{2\pi G M_{i}}{c^{2}R_{i}T_{i}}+\sum_{i=1}^{8}\frac{4\pi GM_{i}r_{i}^{2}}{5c^{2}R_{i}^{3}t_{i}}\nn\\
&=&(2.468\times 10^{-12}-5.266\times 10^{-20}+3.769\times 10^{-25})~{\rm sec^{-1}},
\label{bmsun22}
\eea 
where the first, second and third terms come from the contributions from the spinning effect of the Sun, orbital effects of the solar system 
planets and spinning effects of the solar system planets, respectively. Here $M_{\odot}$, $r_{\odot}$ and $t_{\odot}$ are the mass, 
radius and rotational period of the spinning Sun, while $M_{i}$, $r_{i}$, $R_{i}$, $t_{i}$ and $T_{i}$ $(i=1,2,3,...8)$ are the mass, 
radius, average orbital distance, rotational period  and average orbital period of the $i$-th solar system planet orbiting the Sun, respectively. 
In treating $B_{M}(M_{\odot})$ in (\ref{bmsun22}), we have ignored the effect of the sun radius $r_{\odot}$ which is negligible comparing to 
$R_{Mercury}$ due to the factor $r_{\odot}/R_{Mercury}=1.202\times 10^{-2}$~\cite{carroll96}, for instance, 
in calculating the second term in (\ref{bmsun22}). From now on we will 
ignore the contributions from the spinning effects of the solar system planets in the third term,
since these effects are extremely small, as shown 
in (\ref{bmearth}) for the case of the Earth and Moon system. 

Now we investigate the mass magnetic fields $\vec{B}_{M}(M_{i})=B_{M}(M_{i})\hat{\omega}$ $(i=1,2,3,...,8)$ for the solar system planets 
with $M_{i}$ for an observer located on the surface of the planets where $\hat{\omega}$ is the unit vector along the axis 
direction of the spinning Sun. To do this, using (\ref{bmgma2}) we formulate $B_{M}(M_{i})$ as follows
\beq
B_{M}(M_{i})=\frac{4\pi GM_{i}}{5c^{2}r_{i}t_{i}}+\frac{4\pi GM_{\odot}r_{\odot}^{2}}{5c^{2}R_{i}^{3}t_{\odot}}
+\sum_{j\neq i}\frac{4\pi GM_{j}r_{j}^{2}}{5c^{2}|R_{j}-R_{i}|^{3}t_{j}},
\label{bmmi}
\eeq
where the first, second and third terms originate from the contributions from the spinning effect of the $i$-th planet, spinning 
effect of the Sun and spinning effects of the other $j$-th planets, respectively. Here we 
have made approximation that all the solar system planets are aligned on the straight line in planetary order for simplicity, 
and approximation that for the case of Uranus the direction of $\vec{B}_{M}(M_{7})$ is along the centripetal direction $-\hat{r}$.

Making use of (\ref{bmmi}), for the case of the Earth we arrive at
\beq
B_{M}(M_{\oplus})=(2.029\times 10^{-14}+2.486\times 10^{-19}+2.078\times 10^{-24})~{\rm sec^{-1}},
\label{bmmoplus}
\eeq
where for simplicity we have not included the contribution of $\vec{B}_{M}(M_{7})$ originated from Uranus, since $\vec{B}_{M}(M_{7})$ 
is perpendicular to the directions of $\vec{B}_{M}(M_{i})$ for the other solar system planets. Note that in (\ref{bmmoplus}) 
we can ignore the negligible contributions of the spinning effect of the Sun and 
spinning effects of the other solar system planets. Next in (\ref{bmmoplus}) we also have excluded the negligible contributions of the 
orbital effect of Moon and spinning effect of Moon which are listed in $B_{M}(M_{\oplus})$ in 
(\ref{bmearth}). Moreover, since the contributions of the spinning effects of the other solar system planets with $M_{j}$ $(j=1,2,4,...,8)$ are 
negligible as shown in (\ref{bmmoplus}), in evaluating $B_{M}(M_{i})$ in (\ref{bmmi}) we effectively do not need the 
constraint that all the solar system planets are aligned on the straight line in planetary order.

Next, similar to (\ref{fmbmoon0}) for the case of the Moon, we evaluate the mass magnetic GR force acting on the Earth 
$\vec{F}_{M}^{B}(M_{\oplus}-M_{\odot})=F_{M}^{B}(M_{\oplus}-M_{\odot})\hat{r}$ with 
$\hat{r}$ being the unit radial vector from the Sun. Exploiting (\ref{calftwo2q0}) one can readily 
evaluate $F_{M}^{B}(M_{\oplus}-M_{\odot})$ 
\beq
F_{M}^{B}(M_{\oplus}-M_{\odot})=(3.612\times 10^{15}+4.426\times 10^{10}+3.529\times 10^{5})~{\rm N}\cong 1.019\times 10^{-7}\times 
|F_{M}^{E}(M_{\oplus}-M_{\odot})|,
\label{fmbmopus}
\eeq
where $|F_{M}^{E}(M_{\oplus}-M_{\odot})|$ is given below. Now we readily formulate the mass electric Newtonian force acting on the Earth 
$\vec{F}_{M}^{E}(M_{\oplus}-M_{\odot})=F_{M}^{E}(M_{\oplus}-M_{\odot})\hat{r}$ where $\hat{r}$ is again the unit radial vector from the Sun and 
$F_{M}^{E}(M_{\oplus}-M_{\odot})$ is given by
\bea
F_{M}^{E}(M_{\oplus}-M_{\odot})&=&-\frac{GM_{\oplus}M_{\odot}}{R_{\oplus}^{2}}-\sum_{j=1,2}\frac{GM_{\oplus}M_{j}}{|R_{j}-R_{\oplus}|^{2}}
+\sum_{j>3}\frac{GM_{\oplus}M_{j}}{|R_{j}-R_{\oplus}|^{2}},\nn\\
&=&(-3.543\times 10^{22}-1.149\times 10^{18}+2.103\times 10^{18})~{\rm N},
\label{fmeearth}
\eea
where the first, second and third terms come from the contributions from the Sun, the solar system planets $(i=1,2)$ inside and 
the solar system planets $(i=4,5,...,8)$ outside the Earth orbital radius, respectively. Here 
we again have made approximation that all the solar system planets are aligned on the straight line in planetary order, for simplicity, and approximation that in evaluating $F_{M}^{E}(M_{\oplus}-M_{\odot})$ 
the Sun and the solar system planets do not have the rotational DOF, for simplicity.
Note that the leading order prediction in $F_{M}^{E}(M_{\oplus}-M_{\odot})$ in (\ref{fmeearth}) 
is dominant comparing to the corresponding subleading order ones. Since the contributions of the spinning effects of the other solar system planets 
with $M_{j}$ $(j=1,2,4,...,8)$ are negligible as shown in (\ref{fmeearth}), 
as in the case of the evaluation of $B_{M}(M_{\oplus})$ in (\ref{bmmoplus}), for the leading order prediction 
of $F_{M}^{E}(M_{\oplus}-M_{\odot})$ in (\ref{fmeearth}) we can again remove effectively the constraint that all the solar system planets are aligned on 
the straight line in planetary order.

\begin{table}[t]
\caption{$\vec{B}_{M}(M_{i})=B_{M}(M_{i})\hat{\omega}$, $\vec{F}_{M}^{B}(M_{i}-M_{\odot})=F_{M}^{B}(M_{i}-M_{\odot})\hat{r}$, 
$\vec{F}_{M}^{E}(M_{i}-M_{\odot})=F_{M}^{E}(M_{i}-M_{\odot})\hat{r}$ and $\chi(M_{i}-M_{\odot})=|\vec{F}_{M}^{B}(M_{i}-M_{\odot})|/|\vec{F}_{M}^{E}(M_{i}-M_{\odot})|$ are listed for 
the $i$-th planet $(i=1,2,3,...,8)$. Here $\hat{\omega}$ and $\hat{r}$ are the unit vector along the axis direction 
of the spinning Sun, and that along the radial vector from the Sun, respectively. The magnitudes of $F_{M}^{B}(M_{i}-M_{\odot})$ and 
$F_{M}^{E}(M_{i}-M_{\odot})$ are given in the unit of $|F_{M}^{E}(M_{\oplus}-M_{\odot})|$. 
For the case of Venus, the direction of $\vec{B}_{M}(M_{2})$ is along $-\hat{\omega}$ to produce the centripetal force 
$\vec{F}_{M}^{B}(M_{2}-M_{\odot})$, while for the case of Uranus, the direction of 
$\vec{B}_{M}(M_{7})$ is along $-\hat{r}$ to yield the force $\vec{F}_{M}^{B}(M_{7}-M_{\odot})$ along $\hat{\omega}$.  
The predictions for the exceptional cases of Venus and Uranus are indicated by $*$.}
\begin{center}
\begin{tabular}{lllll}
\hline
$i$-th Planet  &$B_{M}(M_{i})~({\rm sec^{-1}})$ $$&$F_{M}^{B}(M_{i}-M_{\odot})$ 
$$&$F_{M}^{E}(M_{i}-M_{\odot})$ $$&$\chi(M_{i}-M_{\odot})$\\
\hline
Mercury  &$~~$$5.423\times 10^{-17}$ &$~~$$2.422\times 10^{-11}$ &$-3.691\times 10^{-1}$ &$6.563\times 10^{-11}$\\
Venus  &$-7.081\times 10^{-17~*}$ &$-3.409\times 10^{-10~*}$ &$-1.558$ &$2.188\times 10^{-10}$\\
Earth  &$~~$$2.029\times 10^{-14}$ &$~~$$1.019\times 10^{-7}$ &$-1.000$ &$1.019\times 10^{-7}$\\
Mars  &$~~$$3.973\times 10^{-15}$ &$~~$$1.737\times 10^{-9}$ &$-4.622\times 10^{-2}$ &$3.758\times 10^{-8}$\\
Jupiter  &$~~$$1.388\times 10^{-12}$ &$~~$$9.723\times 10^{-4}$ &$-1.174\times 10$ &$8.280\times 10^{-5}$\\
Saturn &$~~$$4.583\times 10^{-13}$ &$~~$$7.097\times 10^{-5}$ &$-1.046$ &$6.784\times 10^{-5}$\\
Uranus  &$~~$$1.021\times 10^{-13~*}$ &$~~$$1.703\times 10^{-6~*}$ &$-3.948\times 10^{-2}$ &$4.314\times 10^{-5}$\\
Neptune  &$~~$$1.355\times 10^{-13}$ &$~~$$2.126\times 10^{-6}$ &$-1.895\times 10^{-2}$ &$1.122\times 10^{-4}$\\
\hline
\end{tabular}
\end{center}
\label{tablechimi}
\end{table}

$\vec{B}_{M}(M_{i})=B_{M}(M_{i})\hat{\omega}$, $\vec{F}_{M}^{B}(M_{i}-M_{\odot})=F_{M}^{B}(M_{i}-M_{\odot})\hat{r}$, 
$\vec{F}_{M}^{E}(M_{i}-M_{\odot})=F_{M}^{E}(M_{i}-M_{\odot})\hat{r}$ and $\chi(M_{i}-M_{\odot})=|\vec{F}_{M}^{B}(M_{i}-M_{\odot})|/|\vec{F}_{M}^{E}(M_{i}-M_{\odot})|$ are listed for the $i$-th planet $(i=1,2,3, \ldots ,8)$. Here, $\hat{\omega}$ and $\hat{r}$ are the unit vector along the axis direction 
of the spinning Sun and that along the radial vector from the Sun, respectively. The magnitudes of $F_{M}^{B}(M_{i}-M_{\odot})$ and 
$F_{M}^{E}(M_{i}-M_{\odot})$ are given in units of $|F_{M}^{E}(M_{\oplus}-M_{\odot})|$. 
For the case of Venus, the direction of $\vec{B}_{M}(M_{2})$ is along $-\hat{\omega}$ to produce the centripetal force 
($\vec{F}_{M}^{B}(M_{2}-M_{\odot})$), while for the case of Uranus, the direction of $\vec{B}_{M}(M_{7})$ is along $-\hat{r}$ to yield the force ($\vec{F}_{M}^{B}(M_{7}-M_{\odot})$) along $\hat{\omega}$. The predictions for the exceptional cases of Venus and Uranus are indicated by $*$.

Next even the subleading order contributions of about order of $10^{18}~{\rm N}$ in $F_{M}^{E}(M_{\oplus}-M_{\odot})$ 
in (\ref{fmeearth}) are greater than the total contribution in $F_{M}^{B}(M_{\oplus}-M_{\odot})$ in (\ref{fmbmopus}), 
implying that $F_{M}^{B}(M_{\oplus}-M_{\odot})$ 
has been hidden in the Newtonian gravity. From (\ref{fmbmopus}) and (\ref{fmeearth}) we arrive at the characteristic 
ratio $\chi(M_{\oplus}-M_{\odot})$ of the form
\beq
\chi(M_{\oplus}-M_{\odot})=|\vec{F}_{M}^{B}(M_{\oplus}-M_{\odot})|/|\vec{F}_{M}^{E}(M_{\oplus}-M_{\odot})|=1.019\times 10^{-7},
\label{ratioofff2}
\eeq
implying that the correction $|\vec{F}_{M}^{B}(M_{\oplus}-M_{\odot})|$ to the Newtonian 
classical prediction $|\vec{F}_{M}^{E}(M_{\oplus}-M_{\odot})|$ is extremely small but it is non-vanishing as in the case of 
(\ref{ratioofff}) for the isolated system of the Earth and the spinning Moon orbiting the spinning Earth.

Now, we investigate the general cases of the solar system planets with masses $M_{i}$ $(i=1,2,3,...,8)$ including the Earth with mass $M_{3}=M_{\oplus}$. 
The predictions of $B_{M}(M_{i})$ 
in (\ref{bmmi}) for the solar system planets with masses $M_{i}$ are given in Table~\ref{tablechimi}. Exploiting (\ref{calftwo2q0}), we can readily construct 
$\vec{F}_{M}^{B}(M_{i}-M_{\odot})=F_{M}^{B}(M_{i}-M_{\odot})\hat{r}$ where $F_{M}^{B}(M_{i}-M_{\odot})$ are listed in 
Table~\ref{tablechimi}.\footnote{In Table 1, the magnitudes of $F_{M}^{B}(M_{i}-M_{\odot})$ and 
$F_{M}^{E}(M_{i}-M_{\odot})$ are given in the unit of $|F_{M}^{E}(M_{\oplus}-M_{\odot})|$ and these values are dimensionless. The 
minus signs in $F_{M}^{E}(M_{i}-M_{\odot})$ again denote the attractive Newtonian forces between $M_{i}$ and $M_{\odot}$.}
Next, in order to find the characteristic ratios, we construct the mass electric Newtonian forces acting on the solar system planets $\vec{F}_{M}^{E}(M_{i}-M_{\odot})=F_{M}^{E}(M_{i}-M_{\odot})\hat{r}$ 
where $F_{M}^{E}(M_{i}-M_{\odot})$ is given by
\beq
F_{M}^{E}(M_{i}-M_{\odot})=-\frac{GM_{i}M_{\odot}}{R_{i}^{2}}-\sum_{j<i}\frac{GM_{i}M_{j}}{|R_{j}-R_{i}|^{2}}
+\sum_{j>i}\frac{GM_{i}M_{j}}{|R_{j}-R_{i}|^{2}}.
\label{fMEMi}
\eeq
Furthermore the predictions for the mass magnetic GR forces $F_{M}^{B}(M_{i}-M_{\odot})$ associated with $B_{M}(M_{i})$ 
in (\ref{bmmi}), the mass electric Newtonian forces $F_{M}^{E}(M_{i}-M_{\odot})$ in (\ref{fMEMi}), and the corresponding characteristic ratios 
\beq
\chi(M_{i}-M_{\odot})=|\vec{F}_{M}^{B}(M_{i}-M_{\odot})|/|\vec{F}_{M}^{E}(M_{i}-M_{\odot})|\sim 10^{-11}-10^{-4}, 
\label{chimimodot}
\eeq
are also listed in Table~\ref{tablechimi}. Here we observe that 
$F_{M}^{B}(M_{i}-M_{\odot})$ and $F_{M}^{E}(M_{i}-M_{\odot})$ acting on the solar system planets with $M_{i}$ show the magnitudes of about order of
$(10^{11}-10^{19})~{\rm N}$ and $(10^{20}-10^{23})~{\rm N}$, respectively. 
Note that for the case of Jupiter 
$B_{M}(M_{5})$, $F_{M}^{B}(M_{5}-M_{\odot})$ and $F_{M}^{E}(M_{5}-M_{\odot})$ have the largest predicted values among those of the solar system planets because of its heaviest mass. 

Note that the smallness of these characteristic ratios $\chi(M_{m}-M_{\oplus})$ and $\chi(M_{i}-M_{\odot})$ originate from the factor $\frac{G}{c^{2}}$ 
involved in $B_{M}(M_{m})$ in (\ref{bmearth2}) and $B_{M}(M_{i})$ in (\ref{bmmi}), with which $F_{M}^{B}(M_{m}-M_{\oplus})$ in (\ref{fmbmoon0}) and 
$F_{M}^{B}(M_{i}-M_{\odot})$ are constructed. In contrast, $F_{M}^{E}(M_{m}-M_{\oplus})$ and $F_{M}^{E}(M_{i}-M_{\odot})$ 
possess the factor $G$ only as shown in (\ref{fMEMoon0}) and (\ref{fMEMi}). 
Note also that, for the case of Venus, the direction of $\vec{B}_{M}(M_{2})$ is along $-\hat{\omega}$ to produce the centripetal force 
$\vec{F}_{M}^{B}(M_{2}-M_{\odot})$. Next, for the case of Uranus, the direction of 
$\vec{B}_{M}(M_{7})$ is along $-\hat{r}$ to yield the force $\vec{F}_{M}^{B}(M_{7}-M_{\odot})$ along $\hat{\omega}$.

\subsection{B-ring of Saturn}

Now it seems appropriate to address some comments on the rings of Saturn, to investigate the corresponding mass magnetic 
dipole moments. It is well known that the B-ring of Saturn is the brightest, most opaque and the most likely massive of Saturn's 
rings~\cite{hedman}. Motivated by this, we study the isolated system of the B-ring of Saturn for simplicity. 
Making use of a multipole expansion valid at distant observation point, we find $\vec{A}_{M}$ in terms of 
the dominant mass magnetic dipole moment contribution\footnote{Here we have exploited the approximation at distant observation point 
$\vec{A}_{M}=-\frac{G}{c^{2}}\left(\frac{1}{r}\oint\vec{J}_{M}d^{3}x^{\prime}
+\frac{x_{i}}{r^{3}}\int\vec{J}_{M}x^{\prime}_{i}d^{3}x^{\prime}
+\cdots\right)$ where the ellipsis stands for the higher order terms which can be ignored for the mass vector 
potential of a localized mass current distribution. Note that the first term denotes the mass magnetic monopole 
term which vanishes for the closed loop integral, as expected. In this paper, we thus exclude the possibility 
of mass magnetic monopole in addition to that of charge magnetic monopole.}
\beq
\vec{A}_{M}=\frac{\vec{m}_{M}\times\vec{x}}{r^{3}},
\label{vecam2}
\eeq
where $r=|\vec{x}|$ and the mass magnetic dipole moment $\vec{m}_{M}$ is given by
\beq
\vec{m}_{M}=-\frac{G}{2c^{2}}\int \vec{x}^{\prime}\times\vec{J}_{M}d^{3}x^{\prime}.
\label{vecmm}
\eeq
Note that the mass magnetic dipole moment $\vec{m}_{M}$ in (\ref{vecmm}) is the analogue of the charge 
magnetic dipole moment $\vec{m}_{Q}$~\cite{jackson,griffiths99em}, and the direction of $\vec{m}_{M}$ is opposite to that of 
$\vec{m}_{Q}$ because of the difference of the interaction patterns between repulsive EM and attractive mass gravitational forces. 

Now we construct the mass magnetic dipole moment of the B-ring orbiting Saturn. To do this, we first find the period $t_{B}$ of the B-ring orbiting Saturn. 
In the absence of non-gravitational fields, the B-ring of Saturn 
moves in circular orbit according to the Kepler's third law to a high degree of approximation~\cite{bastin}, 
where the period $t_{B}$ is 
given by~\cite{goldstein80}
\beq
t_{B}^{2}=\frac{4\pi^{2}R_{B}^{3}}{GM_{6}}.
\label{tb2}
\eeq  
Here $R_{B}$ is the orbital radius of the B-ring. Note that in (\ref{tb2}) we have 
neglected the effect of the B-ring mass $M_{B}$ which is extremely small compared to the mass $M_{6}$. Combining the formulas 
$\vec{m}_{M}$ in (\ref{vecmm}) and $t_{B}$ in (\ref{tb2}), for the B-ring 
of Saturn we finally arrive at the mass magnetic dipole moment $\vec{m}_{M}(Ring)=m_{M}(Ring)\hat{\omega}$ 
where\footnote{The recent Cassini data show that the total 
mass of the rings of Saturn is $1.54\times 10^{19}~{\rm kg}$~\cite{iess19}. In evaluating (\ref{vecmm221}), we have 
exploited this value for the B-ring mass $M_{B}$ for simplicity, since the B-ring is known to be probably the most massive 
of Saturn's rings~\cite{hedman}. Next the orbital radius of the B-ring is given by 
$R_{B}=(9.232\times 10^{7}-1.177\times 10^{8})~{\rm m}$~\cite{carroll96}, and thus we have used the average value 
of $R_{B}=1.050\times 10^{8}~{\rm m}$. Moreover we have ignored the effect of the width of the B-ring of 
Saturn in the evaluation in (\ref{vecmm221}), for simplicity.} 
\beq
m_{M}(Ring)=-\frac{G^{3/2}M_{6}^{1/2}M_{B}R_{B}^{1/2}}{2c^{2}}=-1.141\times 10^{4}~{\rm m^{3}~sec^{-1}}.
\label{vecmm221}
\eeq
Note that the direction of this mass magnetic dipole moment is opposite to that of the charge magnetic dipole moment 
in the electromagnetic interaction, because of the difference of the interaction patterns between charge repulsive electromagnetic 
and mass attractive gravitational forces. Now the predicted value for $m_{M}(Ring)$ in (\ref{vecmm221}) 
is intrinsically associated with the Cassini data~\cite{iess19} concerning the total mass of the rings of Saturn 
via the explicit formula possessing $M_{B}$ in (\ref{vecmm221}), which is the approximated total mass of Saturn's rings~\cite{hedman}. 

Next we investigate the recent Cassini data~\cite{science22} on the formation of 
the young rings through a grazing encounter of an additional satellite, which is named Chrysalis and is assumed to be 
predominantly composed of water ice, with Saturn. This encounter of Chrysalis with Saturn would 
have caused Chrysalis to break apart to yield debris which could have developed into Saturn's young rings~\cite{dones91,science22}. 
The loss of the hypothetical satellite Chrysalis can then explain the obliquity of Saturn and the young age of its rings. Moreover 
the value of $m_{M}(Ring)$ in (\ref{vecmm221}), which could come from the debris mass of Chrysalis via (\ref{vecmm221}), 
thus could be related with the the formation mechanism~\cite{science22} of Saturn's rings.

\section{Conclusions}
\setcounter{equation}{0}
\renewcommand{\theequation}{\arabic{section}.\arabic{equation}}

In summary, we have constructed the formalism of the gravitomagnetism by making use of the MLGR. Exploiting this gravitomagnetism we have evaluated {\it numerically} the astrophysical quantities such as mass magnetic GR force and the mass magnetic dipole moment of the Saturn's B-ring, for instance. 
To be more specific, in order to identify the effectiveness of the gravitomagnetism, we have treated the 
astrophysical observables such as the mass magnetic fields for both the isolated system of the Moon orbiting the spinning Earth and that of the solar system planets orbiting the spinning Sun. Exploiting the mass magnetic fields, 
we have predicted the ensuing Lorentz type mass magnetic GR forces for these systems. 
Note that, in investigating the gravitomagnetism, we have used the MLGR associated with the Wald Approximation, 
different from the LGR possessing the TT gauge. Note also that, in the MLGR related with 
the gravitomagnetism, the angle dependence of the GW radiation can be explained in the vectorial formalism~\cite{hong23}. 
In contrast, the LGR is defined in the second rank tensorial formalism mathematically different from the vectorial one in the 
MLGR, and the LGR can not present the angle dependence of the GW radiation~\cite{hong23}. 
The gravitomagnetism in this paper has predicted the existence and ensuing numerical 
estimates of the physical observable such as the mass magnetic GR force which is hidden in the Newtonian gravity. Note that the LGR can not predict the 
astrophysical physical quantities such as the mass magnetic fields and the mass magnetic GR forces for instance.

Next, exploiting the Maxwell type gravitational equations in the gravitomagnetism we have obtained the mass electric and mass magnetic fields $\vec{E}_{M}$ and $\vec{B}_{M}$ with which we have constructed the 
corresponding mass electric Newtonian and Lorentz type mass magnetic GR forces $\vec{F}_{M}^{E}$ and $\vec{F}_{M}^{B}$. 
One of the advantages of the gravitomagnetism is to evaluate the mass magnetic field $\vec{B}_{M}$, whose phenomenological value is hidden in the Newtonian gravity and is newly predicted in this paper. For instance, in the isolated Earth of mass $M_{\oplus}$ and the Moon of mass $M_{m}$ we have found 
$|\vec{B}_{M}(M_{\oplus})|$ 
acting on the Earth, while in the isolated system of the Sun of mass $M_{\odot}$ and solar system planets of masses $M_{i}$ $(i=1,2,3,...,8)$ we have obtained $|\vec{B}_{M}(M_{\odot})|$ acting on the Sun, which is about $10^{2}\times |\vec{B}_{M}(M_{\oplus})|$. Moreover we have noticed that the mass magnetic GR force $\vec{F}_{M}^{B}$ is relatively small, comparing with $\vec{F}_{M}^{E}$. To be specific, considering the isolated system of the Moon and Earth, 
in the gravitomagnetism we have estimated $|\vec{F}_{M}^{B}(M_{m}-M_{\oplus})|$ acting on the Moon which is approximately $10^{-11}$ of the mass electric Newtonian force $|\vec{F}_{M}^{E}(M_{m}-M_{\oplus})|$. For the case of the Earth and the Sun we have found that $|\vec{F}_{M}^{B}(M_{\oplus}-M_{\odot})|$ is 
about $10^{-7}$ of the corresponding mass electric Newtonian force $|\vec{F}_{M}^{E}(M_{\oplus}-M_{\odot})|$. 
Next for the isolated Sun and solar system planets, in the gravitomagnetism we have evaluated 
$|\vec{F}_{M}^{B}(M_{i}-M_{\odot})|$ acting on the solar system planets, which are about order of $10^{-11}-10^{-4}$ of the corresponding mass electric Newtonian forces $|\vec{F}_{M}^{E}(M_{i}-M_{\odot})|$ as shown in Table~\ref{tablechimi}. This feature implies that the gravitomagnetic correction, namely $|\vec{F}_{M}^{B}(M_{i}-M_{\odot})|$, to the Newtonian classical prediction $|\vec{F}_{M}^{E}(M_{i}-M_{\odot})|$ is 
extremely small but it is non-vanishing. To be specific, this gravitomagnetic correction has been hidden in the astrophysical phenomenology until now. 
The theoretical prediction of the mass magnetic GR force in the gravitomagnetism is one of the main points of this paper. 
Note that $|\vec{F}_{M}^{B}(M_{i}-M_{\odot})|$ possesses the leading order correction, which originates from the spinning effects of 
the solar system planet of $M_{i}$, and this correction is dominant comparing to the other next order ones. 
Note also that the directions of $\vec{F}_{M}^{B}(M_{i}-M_{\odot})$ for the solar system planets except Venus of mass $M_{2}$ and Uranus of mass $M_{7}$ 
are anti-parallel to those of $\vec{F}_{M}^{E}(M_{i}-M_{\odot})$, while the directions of $\vec{F}_{M}^{B}(M_{2}-M_{\odot})$ and $\vec{F}_{M}^{B}(M_{7}-M_{\odot})$ 
for Venus and Uranus are parallel to and perpendicular to those of $\vec{F}_{M}^{E}(M_{i}-M_{\odot})$ $(i=2,7)$, respectively. 

Now, as a typical example of the mass magnetic dipole moment $\vec{m}_{B}$, in the gravitomagnetism we 
have investigated Saturn's B-ring that orbits Saturn, to predict $\vec{m}_{M}(Ring)$. Note that the prediction of $\vec{m}_{M}(Ring)$ has been shown to be closely related with the Cassini data~\cite{iess19} on the total mass of the rings of Saturn. Next $\vec{m}_{M}(Ring)$, which could originate from the debris mass of the hypothetical satellite Chrysalis~\cite{science22}, also could be associated with the formation of the young rings of Saturn. Note that the 
mass magnetic GR force $\vec{F}_{M}^{B}(Ring)$ acting on the B-ring of Saturn contributes in small amounts to the formation of the width of the B-ring. 

Next, we have predicted phenomenologically in this paper the physical quantities such as the mass magnetic fields, their ensuing {\it mass magnetic GR forces} which exist in the Universe as far as we propose the gravitomagnetism (or MLGR). 
It will be interesting to search for observational evidences for these physical quantities 
in the Universe. Once this is done, the theoretical formalism developed in this paper could give some progressive impacts on 
the {\it precision astrophysics phenomenology.} Note that there have been searches on the GR effects such as the relativistic precession 
and frame dragging conducted saliently by the Gravity Probe B satellite that is well documented in Refs.~\cite{everitt15,iorio20gpb,lense18} and references therein.\\ \\

\noindent
{\bf Funding:} This work was supported by Basic Science Research Program through the National Research 
Foundation of Korea funded by the Ministry of Education, Science and Technology RS-2020-NR049598.\\

\noindent
{\bf Data Availability Statement:} No data has been used in the work.\\ 

\noindent
{\bf Acknowledgments:} The author would like to thank the anonymous editor and referees for helpful
comments.\\ 

\noindent
{\bf Conflicts of Interest:} The author declares no conflict of interest.\\ \\

\appendix
\section{Sketch of EM}\label{cemapp}
\setcounter{equation}{0}
\renewcommand{\theequation}{A.\arabic{equation}}

In order to construct the gravitomagnetism formalism discussed in Section 2, we will digress to pedagogically recapitulate the EM associated with charge 
$Q$~\cite{jackson,griffiths99em}. To do this, we start with the charge electric and charge magnetic fields in the EM where we find the Maxwell equations in the EM
\bea
\na\cdot\vec{E}_{Q}&=&\frac{1}{\epsilon_{0}}\rho_{Q},~~~\na\times\vec{E}_{Q}+\frac{\pa\vec{B}_{Q}}{\pa t}=0,\nn\\
\na\cdot\vec{B}_{Q}&=&0,~~~~~~~~\na\times\vec{B}_{Q}-\frac{1}{c^{2}}\frac{\pa\vec{E}_{Q}}{\pa t}=\frac{1}{\epsilon_{0}c^{2}}\vec{J}_{Q},
\label{eqnseight}
\eea
where $\rho_{Q}$ and $\vec{J}_{Q}$ are the volume charge density and volume charge current, respectively. Here we have used 
$\epsilon_{0}\mu_{0}=c^{-2}$ where $\epsilon_{0}$, $\mu_{0}$ and $c$ are the permitivity, permeability and speed of light of free space, respectively.
Note that the charge electric and mass magnetic fields $E_{Q}^{i}$ and $B_{Q}^{i}$ are given in terms of the four charge vector potentials 
$A_{Q}^{\alpha}\equiv\left(\frac{\phi_{Q}}{c},\vec{A}_{Q}\right)$ 
\beq
\vec{E}_{Q}=-\na\phi_{Q}-\frac{\pa \vec{A}_{Q}}{\pa t},~~~
\vec{B}_{Q}=\na\times\vec{A}_{Q}.
\label{linemq}
\eeq
We also find that the charge vector potentials $A_{Q}^{\alpha}$ fulfill 
\beq
\square \phi_{Q}=-\frac{1}{\epsilon_{0}}\rho_{Q},~~~\square \vec{A}_{Q}=-\frac{1}{\epsilon_{0}c^{2}} \vec{J}_{Q},
\label{naphi0q}
\eeq
where we have exploited the Lorentz gauge condition
\beq
\pa_{\alpha}A_{Q}^{\alpha}=0.
\label{pabarh2q}
\eeq
In vacuum, inserting $\rho_{Q}=\vec{J}_{Q}=0$ into (\ref{naphi0q}), we find the wave equation for $A_{Q}^{\alpha}$
\beq
\square A_{Q}^{\alpha}=0,~~~\alpha=0,1,2,3
\label{squareq}
\eeq
which implies a massless spin-one photon propagating in the flat spacetime.

Next we investigate the gauge invariance and U(1) transformation in the EM.
To do this, we start with the Dirac equation for the electron wave function $\psi$ and the charge vector potentials $A_{Q}^{\mu}$ in the unit of $\hbar=c=1$ 
\beq
i\gamma^{\mu}\pa_{\mu}\psi-m_{e}\psi-e\gamma^{\mu}A_{Q,\mu}\psi=0,
\label{diraceqn}
\eeq
where $m_{e}$ and $e$ are the electron mass and charge, respectively. Next the QED describes the interactions between electrons and photons. Now we introduce the QED Lagrangian~\cite{smith86}
\beq
{\cal L}_{QED}=-\frac{1}{4}F_{Q,\mu\nu}F_{Q}^{\mu\nu}+\bar{\psi}(i\gamma^{\mu}\pa_{\mu}-m_{e})\psi-e\bar{\psi}\gamma^{\mu}A_{Q,\mu}\psi,
\label{qcdlag}
\eeq
where $F_{Q}^{\mu\nu}=\pa^{\mu}A_{Q}^{\nu}-\pa^{\nu}A_{Q}^{\mu}$ and the third term denotes the interaction between the electron and the electromagnetic wave (or spin-one photon). Note that the QED Lagrangian is invariant under the gauge transformations
\bea
\psi(x)&\rightarrow& \psi^{\prime}(x)=e^{-ie\Gamma(x)}\psi(x),\nn\\
A_{Q}^{\mu}(x)&\rightarrow& A_{Q}^{\prime\mu}(x)=A_{Q}^{\mu}(x)+\pa^{\mu}\Gamma(x),
\label{trfm22}
\eea
where the first relation is the U(1) transformation.

Next, for the cases of charge electrostatics and charge magnetostatics, corresponding to $\rho_{Q}$ and $\vec{J}_{Q}$ 
inside a given volume we obtain the charge electric and charge magnetic fields
\beq
\vec{E}_{Q}=\frac{1}{4\pi\epsilon_{0}}\int\frac{\rho_{Q}\hat{R}}{R^{2}}d^{3}x^{\prime},~~~
\vec{B}_{Q}=\frac{1}{4\pi \epsilon_{0}c^{2}}\int\frac{\vec{J}_{Q}\times\hat{R}}{R^{2}}d^{3}x^{\prime},
\label{chemfields}
\eeq
where $\vec{R}=\vec{x}-\vec{x}^{\prime}$ is the vector from $d^{3}x^{\prime}$ to the field point $\vec{x}$ and 
$\hat{R}=\vec{R}/R$ with $R=|\vec{x}-\vec{x}^{\prime}|$. Exploiting the charge electric and magnetic fields $\vec{E}_{Q}$ and $\vec{B}_{Q}$ 
in (\ref{chemfields}) we end up with the corresponding force acting on a test charge $Q_{0}$ moving with velocity $v$
\beq
\vec{F}_{Q}=Q_{0}(\vec{E}_{Q}+\vec{v}\times\vec{B}_{Q}).
\label{calftwo2q}
\eeq
Moreover, in the charge magnetostatics, we construct the charge vector potential $\vec{A}_{Q}$ as follows
\beq
\vec{A}_{Q}=\frac{1}{4\pi\epsilon_{0}c^{2}}\int \frac{\vec{J}_{Q}}{R}d^{3}x^{\prime}.
\label{vecam0}
\eeq
Making use of a multipole expansion valid at distant observation point, we reformulate $\vec{A}_{Q}$ in terms of 
the dominant charge magnetic dipole moment contribution
\beq
\vec{A}_{Q}=\frac{\vec{m}_{Q}\times\vec{x}}{r^{3}},
\label{vecaq}
\eeq
where $r=|\vec{x}|$ and the charge magnetic dipole moment $\vec{m}_{Q}$ is given by
\beq
\vec{m}_{Q}=\frac{1}{8\pi\epsilon_{0}c^{2}}\int \vec{x}^{\prime}\times\vec{J}_{Q}d^{3}x^{\prime}.
\label{vecmq}
\eeq



\begin{thebibliography}{99}

\bibitem{ramirez18} Deriglazov, A.A.; Ramirez, W.G. Frame-dragging effect in the field of non rotating body due to unit gravimagnetic moment. {\it Phys. Lett. B} {\bf 2018}, 779, 210-213.
\bibitem{everitt15} Everitt, C.W.F; Muhlfelder, B.; DeBra, D.B.; Parkinson, B.W.; Turneaure, J.P.;
Silbergleit, A.S.; Acworth, E.B.; Adams, M.; Adler, R.; Bencze, W.J.; et al. The Gravity Probe B test of general relativity. 
{\it Class. Quant.Grav.} {\bf 2015}, 32, 224001.

\bibitem{lense18} Lense, J.; Thirring, H. Ueber den Einfluss der Eigenrotation der Zentralkoerper auf 
die Bewegung der Planeten und Monde nach der Einsteinschen Gravitationstheorie. {\it Phys. Z.} {\bf 1918}, 19, 156. 
English translation of this paper appears in Mashhoon, B.; Hehl, F.W.; Theiss, D.S. {\it Gen. Rel. Grav.} {\bf 1984}, 16, 711.
\bibitem{iorio09} Iorio, L. Will the recently approved LARES mission be able to measure the Lense-Thirring effect at 1\%?. {\it Gen. Rel. Grav.} {\bf 2009}, 41, 1717-1724.

\bibitem{hobson06} Hobson, M.P.; Efstathiou, G.; Lasenby, A.N.  {\it General Relativity: 
An Introduction for Physicists}; Cambridge University Press: Cambridge, UK 2006.
\bibitem{ludwig21} Ludwig. G.O. Galactic rotation curve and dark matter according to gravitomagnetism. 
{\it Eur. Phys. J. C} {\bf 2021}, 81, 186.
\bibitem{speake22} Ruggiero, M.L.; Ortolan,  A.; Speake, C.C. Galactic dynamics in general relativity: 
the role of gravitomagnetism. {\it Class. Quant.Grav.} {\bf 2022}, 39, 225015.
\bibitem{jones23} Glampedakis, K.; Jones, D.I. Pitfalls in applying gravitomagnetism to galactic 
rotation curve modelling. {\it Class. Quant. Grav.} {\bf 2023}, 40, 147001.
\bibitem{lasenby23} Lasenby, A.N.; Hobson, M.P.; Barker, W.E.V. Gravitomagnetism and galaxy rotation curves: a cautionary tale. {\it Class. Quant. Grav.} {\bf 2023}, 40, 215014.


\bibitem{hong23} Hong, S.T. New algorithm of measuring gravitational wave radiation from rotating binary system. {\it Phys. Scripta} {\bf 2024}, 99, 105030. 
\bibitem{wald84} Wald, R.M. {\it General Relativity}; University of Chicago Press: Chicago, IL, USA, 1984.
\bibitem{landau75} Landau, L.D.; Lifschits, E.M. {\it The Classical Theory of Fields}; Pergamon Press: Oxford, UK, 1975.

\bibitem{ward04} Ward, W.R.; Hamilton, D.P. Tilting Saturn. I. analytic model. {\it Astrophys. J.} {\bf 2004}, 128, 2501.
\bibitem{goldreich82} Goldreich, P.; Tremaine, S. The dynamics of planetary rings. {\it Ann. Rev. Astron. Astrophys.} {\bf 1982}, 20, 249-283.

\bibitem{zhang17} Zhang, Z.; Hayes, A.G.; Janssen, M.A.; Nicholson, P.D.; Cuzzi, J.N.; de Pater, I.; Dunn, D.E. 
Exposure age of Saturn's A and B rings, and the Cassini Division as suggested by their non-icy material content. 
{\it Icarus} {\bf 2017}, 294, 14-42.

\bibitem{iess19} Iess, L.; Militzer B.; Kaspi, Y.; Nicholson, P.; Durante, D.; Racioppa, P.; Anabtawi, A.; Galanti, E.; Hubbard, W.; Mariani, M. J.; et al. Measurement and implications of Saturn's gravity field and ring mass. {\it Science} {\bf 2019}, 364, 2965.

\bibitem{science22} Wisdom, J.; Dbouk, R.; Militzer,B.; Hubbard, W.B.; Nimmo,F.; Downey, B.G.; French, R.G. Loss of a satellite could explain Saturn's obliquity and young rings. {\it Science} {\bf 2022}, 377. 1285.
\bibitem{jackson} Jackson, J.D. {\it Classical Electrodynamics}; John Wiley \& Sons: Danvers, MA, USA, 1999.
\bibitem{griffiths99em} Griffiths, D. {\it Introduction to Electrodynamics}, Prentice Hall: Upper Saddle River, NJ, USA, 1999.

\bibitem{carroll96} Carroll, B.W.; Ostlie, D.A. {\it Modern Astrophysics}; Addison-Wesley: New York, NY, USA, 1996.

\bibitem{smith86} De Wit, B.; Smith, J. {\it Field Theory in Particle Physics}, North-Holland: Amsterdam, Netherlands, 1986.
\bibitem{pdg} Navas, S.; Amsler, C.; Gutsche, T.; Hanhart, C.; Hern\'andez-Rey J.J.; Louren\c{c}o, C.; Masoni, A.; Mikhasenko, 
M.; Mitchell, R.E.; Patrignani, C.; et al. (Particle Data Group). Review of particle physics. {\it Phys. Rev. D} {\bf 2024}, 110, 030001.
\bibitem{padman} Padmanabhan, T. {\it Theoretical Astrophysics}, Vol. II, Cambridge University Press: Cambridge, UK, 2001.
\bibitem{hedman} Hedman, M.M.; Nicholson, P.D. The B-rings's surface mass density from hidden density waves: Less than meets 
the eye?. {\it Icarus} {\bf 2016}, 279, 109.
\bibitem{bastin} Bastin, J.A. Note on the rings of Saturn. {\it The Moon and the Planets} {\bf 1981}, 24, 467.
\bibitem{goldstein80} Goldstein, H. {\it Classical Mechanics}; Addison-Wesley: Boston, MA, USA, 1980.
\bibitem{dones91} Dones, L. A recent cometary origin for Saturn's rings?. {\it Icarus} {\bf 1991}, 92, 194-203.
\bibitem{iorio20gpb} Iorio, L. Is there still something left that Gravity Probe B can measure?. {\it Universe} {\bf 2020}, 6, 85.


\end{thebibliography}
\end{document}